\documentclass[12pt]{iopart}

\usepackage{graphicx}

\begin{document}

\title[Electric fields of the filamentation instability]{Electric field 
generation by the electron beam filamentation instability: Filament size 
effects}

\author{M E Dieckmann\textsuperscript{1} and A Bret\textsuperscript{2}}
\address{1 VITA, Department of Science and Technology (ITN), Link\"oping 
University, Campus Norrk\"oping, 60174 Norrk\"oping, Sweden}
\address{2 ETSI Industriales, Universidad de Castilla-La Mancha, 13071 Ciudad
Real, Spain}
\ead{Mark.E.Dieckmann@itn.liu.se}

\begin{abstract}
The filamentation instability (FI) of counter-propagating beams of electrons 
is modelled with a particle-in-cell simulation in one spatial dimension and 
with a high statistical plasma representation. The simulation direction is 
orthogonal to the beam velocity vector. Both electron beams have initially
equal densities, temperatures and moduli of their nonrelativistic mean 
velocities. The FI is electromagnetic in this case. A previous study of a
small filament demonstrated, that the magnetic pressure gradient force (MPGF) 
results in a nonlinearly driven electrostatic field. The probably small 
contribution of the thermal pressure gradient to the force balance implied, 
that the electrostatic field performed undamped oscillations around a 
background electric field. Here we consider larger filaments, which reach a 
stronger electrostatic potential when they saturate. The electron heating 
is enhanced and electrostatic electron phase space holes form. The competition 
of several smaller filaments, which grow simultaneously with the large 
filament, also perturbs the balance between the electrostatic and magnetic 
fields. The oscillations are damped but the final electric field amplitude is 
still determined by the MPGF.  
\end{abstract}

\pacs{52.35.Qz,52.35.Mw,52.65.Rr}

\maketitle

\section{Introduction}
The filamentation instability (FI) driven by counterpropagating electron
beams amplifies magnetic fields in astrophysical and solar flare plasmas 
[1-5] and it is also relevant for inertial confinement fusion (ICF) 
\cite{Tabak} and laser-plasma interactions in general \cite{Ruhl,Borghesi}. 
It has been modelled with particle-in-cell (PIC) and Vlasov codes [9-17] 
taking sometimes into account the ion response and a guiding magnetic field. 
It turns out that the FI is important, when the beam speeds are at least 
mildly relativistic and if the beams have a similar density \cite{Bret}. 
Otherwise its linear growth rate decreases below those of the competing 
two-stream instability or mixed mode instability \cite{Bret2}.

The saturation of the FI is attributed to magnetic trapping \cite{Davidson}.
More recently, it has been pointed out \cite{Pukhov2,Califano} that the 
electric fields are also important in this context. An electric field 
component along the beam velocity vector $\bi{v}_b$ is driven by the FI 
through the displacement current. This component is typically weak and its 
relevance to the plasma dynamics is negligible compared to that of the 
magnetic and the electrostatic fields. The FI is partially electrostatic 
during its linear growth phase, if the electron beams are asymmetric due
to different densities. Symmetric electron beams result in purely 
electromagnetic waves with wavevectors $\bi{k} \perp \bi{v}_b$ 
\cite{Bret2,Tzoufras}. A nonlinear growth mechanism is provided in this 
case by the current of the electrons, which have been accelerated by the 
magnetic pressure gradient force (MPGF). 

The electromagnetic and electrostatic components separate in a 1D simulation 
box, because the gradients along two directions vanish in the Maxwell's 
equations. The electrostatic field is polarized in the simulation direction, 
while the electromagnetic components are polarized orthogonal to it. If both 
electron beams have an equal density and temperature, the electrostatic field 
component along the wavevector $\bi{k}$ can only be driven nonlinearly.
We select here a direction of our 1D PIC simulation box that is orthogonal 
to $\bi{v}_b$, through which this nonlinear mechanism can be examined in an 
isolated form. The equally dense and warm counterstreaming beams of electrons 
have the velocity modulus $|\bi{v}_b| = 0.3c$. The ions are immobile and
compensate the electron charge. The mildly relativistic relative streaming 
speed $\approx 0.55c$ implies, that the growth rate of the FI is significant. 
At the same time, any relativistic mass changes can be neglected during the 
growth phase and the saturation of the FI.

The initial conditions of the plasma equal those in the Refs. 
\cite{Rowlands,NewPoP}. The size distribution of the filaments could be 
sampled with the help of the long 1D simulation box in Ref. \cite{Rowlands}. 
A pair of current filaments, which are small according to this size 
distribution, has been isolated in Ref. \cite{NewPoP}. It could be shown that 
the electrostatic field is indeed driven by the MPGF for this filament pair. 
The electrostatic field performed undamped oscillations around a background 
one. The latter excerted the same force on the electrons as the MPGF. Here 
we assess the influence of the filament size. 

This paper is structured as follows. Section 2 discusses briefly the PIC 
code, the initial conditions and the key nonlinear processes. The results 
are presented in the section 3, which can be summarized as follows. The 
electrons are heated up along the wavevector $\bi{k}$ by their interaction 
with the wave fields. As we increase the filament size the peak amplitudes
grow, which are reached by the magnetic and by the electrostatic field when 
the FI saturates. The electron heating increases with the filament size and
large electron phase space holes form, which interact with the electromagnetic
fields of the filamentation modes. The large box sizes allow the growth of 
more than one wave and the filamentation modes compete. The electrostatic 
field oscillations are damped or inhibited and the amplitude modulus 
converges to one, which equals that expected from the MPGF. We confirm that 
the strength of the electrostatic force on an electron is comparable to that 
of the magnetic force, when the FI saturates. The extraordinary modes are 
pumped by the FI \cite{Califano}. The results are discussed in section 4.

\section{The PIC simulation, the initial conditions and the nonlinear terms}

The PIC simulation method is detailed in Ref. \cite{Dawson}. Our code is 
based on the numerical scheme proposed by \cite{Eastwood}. The phase space 
fluid is approximated by an ensemble of computational particles (CPs) with 
a mass $m_{cp}$ and charge $q_{cp}$ that can differ from those of the 
represented physical particles. The charge-to-mass ratio must be preserved 
though. The Maxwell-Lorentz equations are solved. The plasma frequency of 
each beam with the density $n_e$ that we model is $\omega_p = {(e^2n_e/m_e 
\epsilon_0)}^{0.5}$ and $\Omega_p = \sqrt{2}\omega_p$. The electric and 
magnetic fields are normalized to $\bi{E}_N = e\bi{E}/ c m_e \Omega_p$ and 
$\bi{B}_N = e\bi{B} /m_e \Omega_p$. The current is normalized to $\bi{J}_N 
= \bi{J} / 2 n_e e c$ and the charge to $\rho_N = \rho / 2 n_e e$. The 
physical position, the time and speed are normalized as $x_N = x/ \lambda_s$ 
with $\lambda_s = c / \Omega_p$, $t_N = t \Omega_p$ and $\bi{v}_N = \bi{v}/c$.
The normalized frequency $\omega_N = \omega / \Omega_p$. We drop the indices 
$N$ and $x,t,\omega,\bi{E},\bi{B}, \bi{J}$ and $\rho$ are specified in 
normalized units. The equations are
\begin{eqnarray}
\nabla \times \bi{E} = -\partial_t \bi{B} \, , 
\, \, \nabla \times \bi{B} = \bi{J} + \partial_t \bi{E}, \\
\nabla \cdot \bi{E} = \rho, \, \, \, \nabla \cdot \bi{B} = 0, \\
\rmd_t \bi{p}_{cp} = q_{cp} \left ( \bi{E}[x_{cp}] 
+ \bi{v}_{cp} \times \bi{B}[x_{cp}] \right ) \, , \, \, 
\rmd_x x_{cp} = v_{cp,x},
\end{eqnarray}
with $\bi{p}_{cp} = m_{cp} \Gamma_{cp} \bi{v}_{cp}$. Here $v_{cp,x}$ is 
the component along $x$ of $\bi{v}_{cp}$. The currents $\bi{j}_{cp} 
\propto q_{cp} \bi{v}_{cp}$ of each CP are interpolated to the grid. The 
summation over all CPs gives $\bi{J}$, which is defined on the grid. The 
$\bi{J}$ updates $\bi{E}$ and $\bi{B}$ through (1). Our numerical scheme 
fulfills (2) as constraints. The new fields are interpolated to the position 
of each CP and advance its position $x_{cp}$ and $\bi{p}_{cp}$ through (3). 
All components of $\bi{p}$ are resolved.

\begin{figure}
\begin{center}
\includegraphics[width=12.5cm]{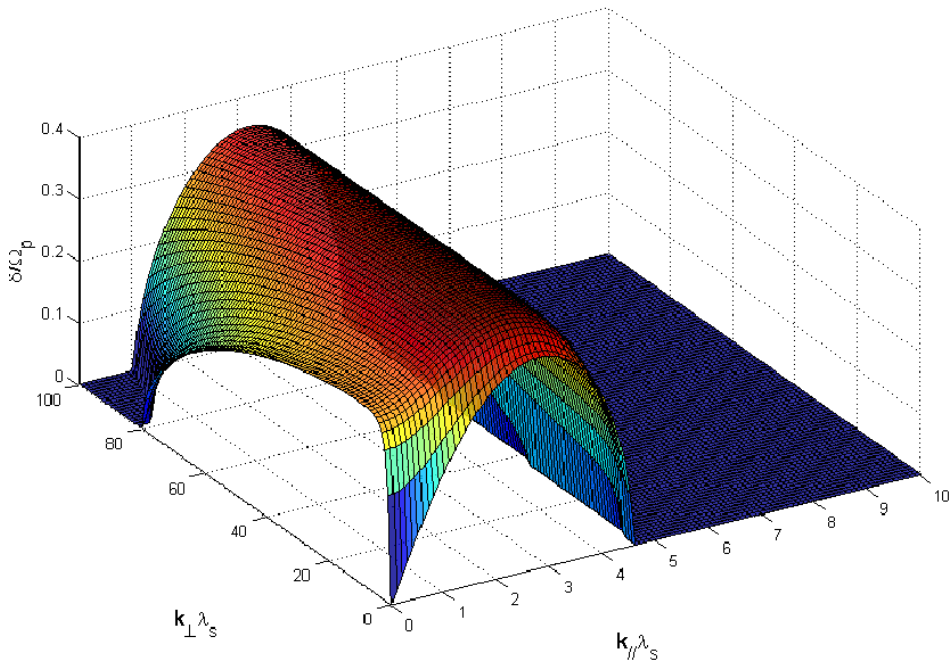}
\caption{(Colour online) The growth rates in units of $\Omega_p$ as a 
function of the wavenumber in the full $\bi{k}$ space, where $\lambda_s 
k_\parallel$ ($\lambda_s k_\perp$) points along (orthogonal) to $\bi{v}_b$. 
The growth rates of the FI modes with $k_\parallel = 0$ are comparable to that 
of the two-stream mode with $k_\perp = 0$ and to those of the oblique modes. 
The growth rates for $k_\parallel = 0$ decrease to zero for $k_\perp 
\rightarrow 0$ and they are stabilized at high $k_\perp$ by thermal effects. 
The growth rate maximum for $k_\parallel = 0$ is $\delta_M = 0.29$ and it is 
reached at $k_M \lambda_s \approx 10$.}\label{Mfg1}
\end{center}
\end{figure}

Two spatially uniform beams of electrons with $q_{cp}/m_{cp}=-e/m_e$ move 
along $z$. Beam 1 has the mean speed $v_{b1} = v_b$ and the beam 2 has 
$v_{b2}=-v_{b1}$ with $v_b = 0.3$. Both beams have a Maxwellian velocity 
distribution in their respective rest frame with a thermal speed $v_{th}= 
c^{-1} {(k_b T / m_e)}^{0.5}$ of $v_b / v_{th} = 18$. The negative electron 
charge is compensated by an immobile positive charge background. The initial 
conditions are $\rho, \bi{J}, \bi{E}, \bi{B}=0$. Figure \ref{Mfg1} displays 
the $\bi{k}$ spectrum of the unstable waves.
The growth rates of the FI modes are close to the maximum value, while
relativistic effects are still negligible. The growth rate spectrum with 
$k_\parallel = 0$ relevant for our simulations peaks with $\delta_M = 0.29$ 
at $k_M \lambda_s \approx 10$. A filamentation mode with $k_M \lambda_s =7$ 
has been considered in detail previously \cite{NewPoP}, while we investigate 
here larger filaments. The box length $L_1 = 2$ for the simulation 1 and the 
filamentation mode with $k_1 = 2\pi / L_1$ grows at the exponential rate 
$0.92 \, \delta_M$. The box length of the simulation 2 is $L_2=2.8$ and the 
growth rate of the filamentation mode with $k_2= 2\pi /L_2$ is $0.86 \, 
\delta_M$. The growth rates decrease rapidly for lower $k$ and these modes
are no longer observed in PIC simulations \cite{Rowlands}. Both simulations 
resolve $x$ by $N_g = 500$ grid cells with the length $\Delta_x$ and use 
periodic boundary conditions. The phase space distributions $f_1(x,\bi{v})$ 
of beam 1 and $f_2(x,\bi{v})$ of beam 2 are each sampled by $N_p = 6.05 
\cdot 10^7$ CPs. The total phase space density is defined as 
$f(x,\bi{v})=f_1(x,\bi{v})+f_2(x,\bi{v})$. 

Each electron beam constitutes prior to the saturation of the FI a fluid 
with the index $j$, which has the density $n_j (x) = \int_{\bi{v}} 
f_j(x,\bi{v}) d\bi{v}$ and the mean velocity $\bi{v}_j (x) = \int_{\bi{v}} 
\bi{v} f_j(x,\bi{v}) d\bi{v}$. The normalized momentum equation for such a 
fluid is 
\begin{equation}
\partial_t (n_j\bi{v}_j) + \nabla (n_j\bi{v}_j\bi{v}_j) = -\nabla \mathbf{P}_j
- n_j\bi{E} + \nabla (\bi{B}\bi{B}) - \nabla \bi{B}^2/2 +\bi{B} \times 
\partial_t \bi{E},
\end{equation}
where the thermal pressure tensor $\mathbf{P}_j$ is normalized to $2 m_e
n_ec^2$. The restriction to one spatial dimension implies, that the 
gradients along $y$ and $z$ vanish. The FI results in this case in the 
initial growth of $B_y$ and of a weaker electric $E_z$. The thermal pressure 
is initially diagonal due to the spatially uniform single-Maxwellian velocity 
distribution. The x-component of the simplified fluid momentum equation is
\begin{equation}
\partial_t (n_j v_{j,x}) + \rmd_x (n_j v_{j,x}^2) = -v_{th}^2 \rmd_x n_j 
-n_j E_x - B_y \rmd_x B_y + B_y \partial_t E_z.\label{1D}
\end{equation}
The thermal pressure gradient $v_{th}^2 \rmd_x n_j$ is valid, as long 
as the electron beams have not been heated up. Let us assume that the 
displacement current and the thermal pressure gradient can be neglected, 
leaving us with the term $n_j E_x$ and the MPGF as the key nonlinear 
terms. The fluid momentum equations can be summed over both beams and we 
consider the right hand side of (\ref{1D}). As long as $E_x$ is small, the 
electron density is not spatially modulated and $n_1+n_2 \approx 1$. The 
nonlinear terms cancel out, if $E_x = -2B_y \rmd_x B_y$. It could be 
demonstrated for a short filament in Ref. \cite{NewPoP} that this is 
the case, even when the FI just saturated. The $E_x$ oscillated in time and 
after the saturation with the amplitude $E_B = -B_y \rmd_x B_y$ around a 
time-stationary $E_B$.

\section{Simulation results}

\subsection{The scaling of $B_y$, $E_x$ and $E_B$ with the box length}
 
The beam velocity $\bi{v}_b \parallel \bi{z}$ and the electrons of both 
beams and their micro-currents are re-distributed by the FI only along $x$. 
The initially charge- and current-neutral plasma is transformed into one 
with $J_z (x,t) \neq 0$. The gradients along the $y,z$-direction vanish in 
our 1D geometry. Ampere's law simplifies to $d_x B_y = J_z + \partial_t E_z$,
resulting in the growth of $B_y$ and $E_z$. The MPGF drives $E_x$. The 
$B_x = 0$ in the 1D geometry and $E_y,B_z$ remain at noise levels. The 
right-hand side of (\ref{1D}) depends on $E_x$, $E_z$ and $B_y$, as well 
as on their spatial gradients, which should vary with the filament size. 

\begin{figure}
\begin{center}
\includegraphics[width=0.49\textwidth]{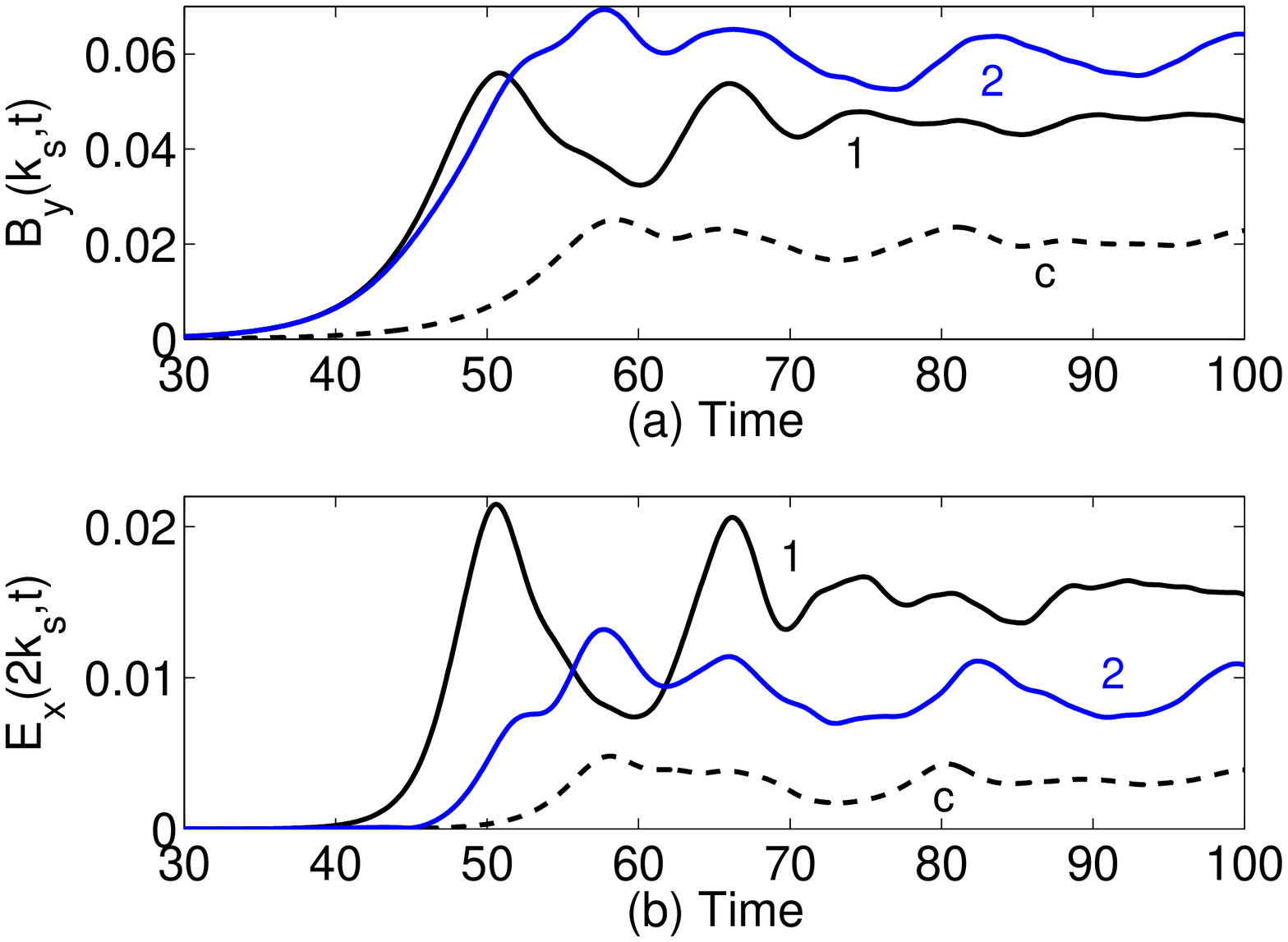}
\includegraphics[width=0.49\textwidth]{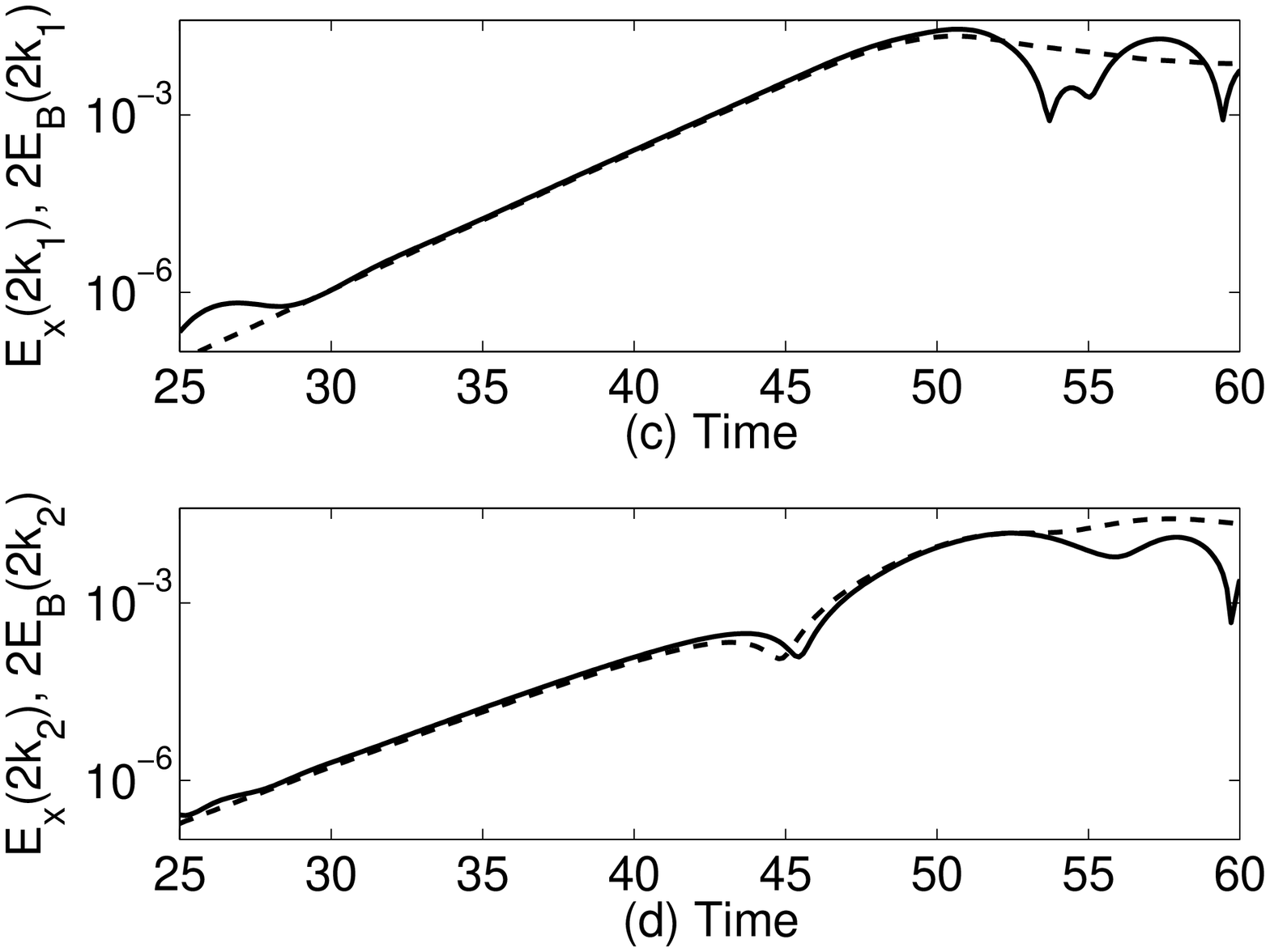}
\caption{(Colour online) Panel (a) compares the $B_y (k_s,t)$ and panel (b) 
the $E_x(2k_s,t)$ in the boxes with the size $L_1$, $L_2$ and $L_c$ (dashed 
curve). $E_B (2k_s,t)$ (dashed curve) is compared with $E_x (2k_s,t)$ (solid 
curves) for the box size $L_1$ (c) and $L_2$ (d).}\label{Mfg2}
\end{center}
\end{figure}

We want to gain qualitative insight into the scaling of the field amplitudes 
with the filament size and determine if $E_x$ is driven by the MPGF 
also for the large filaments. The fields that grow in simulation 1 and 2 are 
compared to those discussed previously in Ref. \cite{NewPoP} that used the 
box size $L_c = 0.89$. Figure \ref{Mfg2} shows the respective dominant 
Fourier component of $B_y$, of $E_x$ and of $2E_B$. The amplitude moduli of 
the mode with $k_s = 2\pi / L_s$ are considered for $B_y$ and those of the 
$2k_s$ mode for $E_x$ and $2E_B$. The subscript $s$ is 1, 2 or $c$ and 
refers to the respective simulation.
The amplitudes of $B_y$ increase with an increasing box size. After the
FI has saturated, we find that $B_y(k_1,t) \approx 2 B_y(k_c,t)$ and 
$B_y(k_2,t) \approx 2.5 B_y (k_c,t)$. The increase of the saturation value 
of $B_y (k_s,t)$ with $L_s$ is consistent with magnetic trapping 
\cite{Davidson}. The magnetic bouncing frequency $\omega_b = {(v_b k_s 
B[k_s,t])}^{1/2}$ in our normalization. The FI should saturate once 
$\omega_b$ is comparable to the linear growth rate of the FI, which is 
approximately constant for the box sizes $L_c$, $L_1$ and $L_2$ (Fig. 
\ref{Mfg1}). A lower $k_s$ supports a larger $B_y(k_s,t)$. The $\omega_b 
\approx 0.2$ for simulation 1 is comparable to the linear growth rate 
$\omega_i \approx 0.25$.

After the saturation, the $E_x(2k_{1,2},t)> 2 E_x(2k_c,t)$ and $E_x(2k_1,t)
> E_x(2k_2,t)$. The $E_x (2k_1,t) > 3 E_x(2k_c,t)$ while $L_1/L_c \approx 
2.2$. The electrostatic potential in simulation 1 is thus larger by a 
factor 6, which should result in a more violent electron acceleration than 
in the box with the length $L_c$. The thermal pressure gradient force is
potentially more important for larger filaments and it may modify the 
balance between the nonlinearly driven $E_x$ and the MPGF. However, an 
excellent match between $E_x(2k_1,t)$ and $2E_B(2k_1,t)$ is observed for 
$t < 50$, due to which the two nonlinear terms on the right hand side of 
(\ref{1D}) practically cancel for simulation 1. The $E_x(2k_2,t)\approx 
2E_B(2k_2,t)$ in simulation 2 for $30 < t < 42$ and for $46 < t < 53$. 
Both fields disagree in between these time intervals and a local minimum 
is observed. The field and electron dynamics is now examined in more 
detail for the box lengths $L_1$ and $L_2$.  

\subsection{Simulation 1: Box length $L_1 = 2$}

\begin{figure}
\begin{center}
\includegraphics[width=0.49\columnwidth]{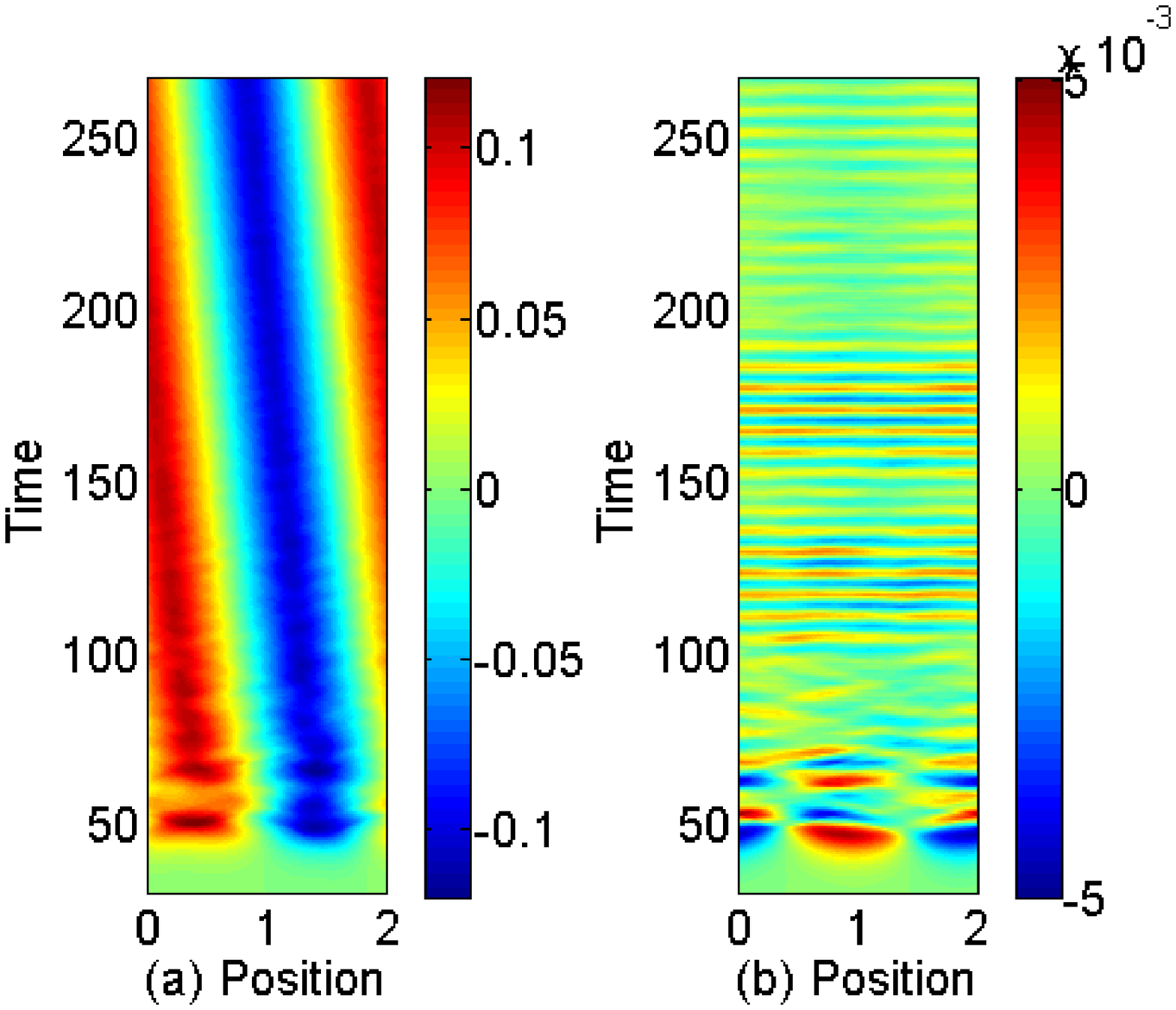}
\includegraphics[width=0.49\columnwidth]{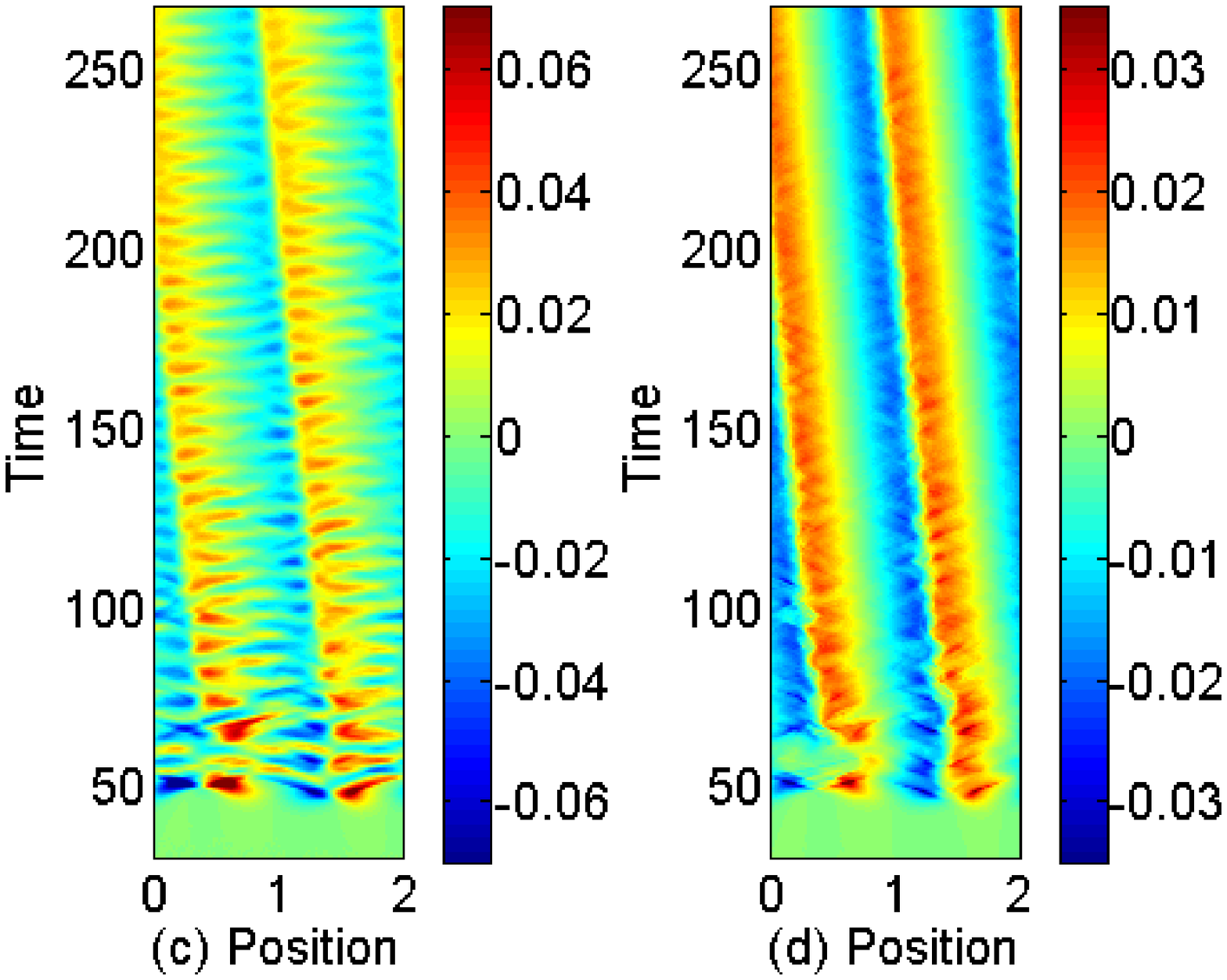}
\caption{(Colour online) The field amplitudes in the box $L_1$: The 
panels (a-d) show $B_y$, $E_z$, $E_x$ and $E_B$, respectively. The 
amplitude of $B_y$ reaches a time-stationary distribution, which 
convects to decreasing $x$ at a speed $<0.01$. The $E_z$ and 
$E_x$ components are oscillatory in space and in time. The $E_z$ 
is phase-shifted by $90^\circ$ relative to $B_y$ when the fields 
saturate at $t\approx 45$. The $E_x$ and the $E_B$ are co-moving 
and $E_x$ oscillates in time around a mean amplitude comparable to 
$E_B$ for $t > 70$.\label{Mfg3}}
\end{center}
\end{figure}

Figure \ref{Mfg3} displays the evolution of the relevant field components.
The $B_y(x,t)$ rapidly grows and saturates at $t\approx 45$. It is initially 
stationary in space but it oscillates in time until $t\approx 65$, which 
implies that $B_y (x,t)$ does not immediately go into its stable saturated 
state. The $B_y(x,t)$ shows only one spatial oscillation and the 
filamentation mode with the wavelength $k_1= 2\pi / L_1$ is thus strongest. 
However, the interval with the large positive $B_y (x,t \approx 45)$ covers 
$0 < x < 0.9$, while that with the large negative $B_y (x,t\approx 45)$ is 
limited to $1.2 < x < 1.7$. This mode is thus initially not monochromatic.
The saturated structure formed by $B_y(x,t)$ drifts after $t\approx 65$ to 
lower $x$ at a speed $<0.01$ and it remains stationary in its moving rest 
frame. The $E_z(x,t)$ grows initially in unison with $B_y (x,t)$ and it is 
shifted in space by $90^\circ$ with respect to $B_y (x,t)$, as expected 
from Ampere's law. Oscillations of $E_z (x,t)$ are spatially correlated with
those of the $B_y(x,t)$ for $45 < t < 65$. The $E_z (x,t)$ undergoes a mode 
conversion at $t\approx 65$ into a time-oscillatory and spatially uniform 
$E_z(x,t)$. Figure \ref{Mfg3}(c) demonstrates that $E_x (x,t)$ is following 
the drift of $B_y(x,t)$ towards decreasing $x$, but that its wavenumber is 
twice that of $B_y(x,t)$. The $B_y (x,t)$ is stationary in its moving rest 
frame, while $E_x(x,t>70)$ is oscillating around an equilibrium electric 
field with an amplitude and spatial distribution that resembles $E_B(x,t)$ 
in Fig. \ref{Mfg3}(d). The electric and the magnetic forces are comparable
in their strength, but their distribution differs.

\begin{figure}
\begin{center}
\includegraphics[width=0.45\columnwidth]{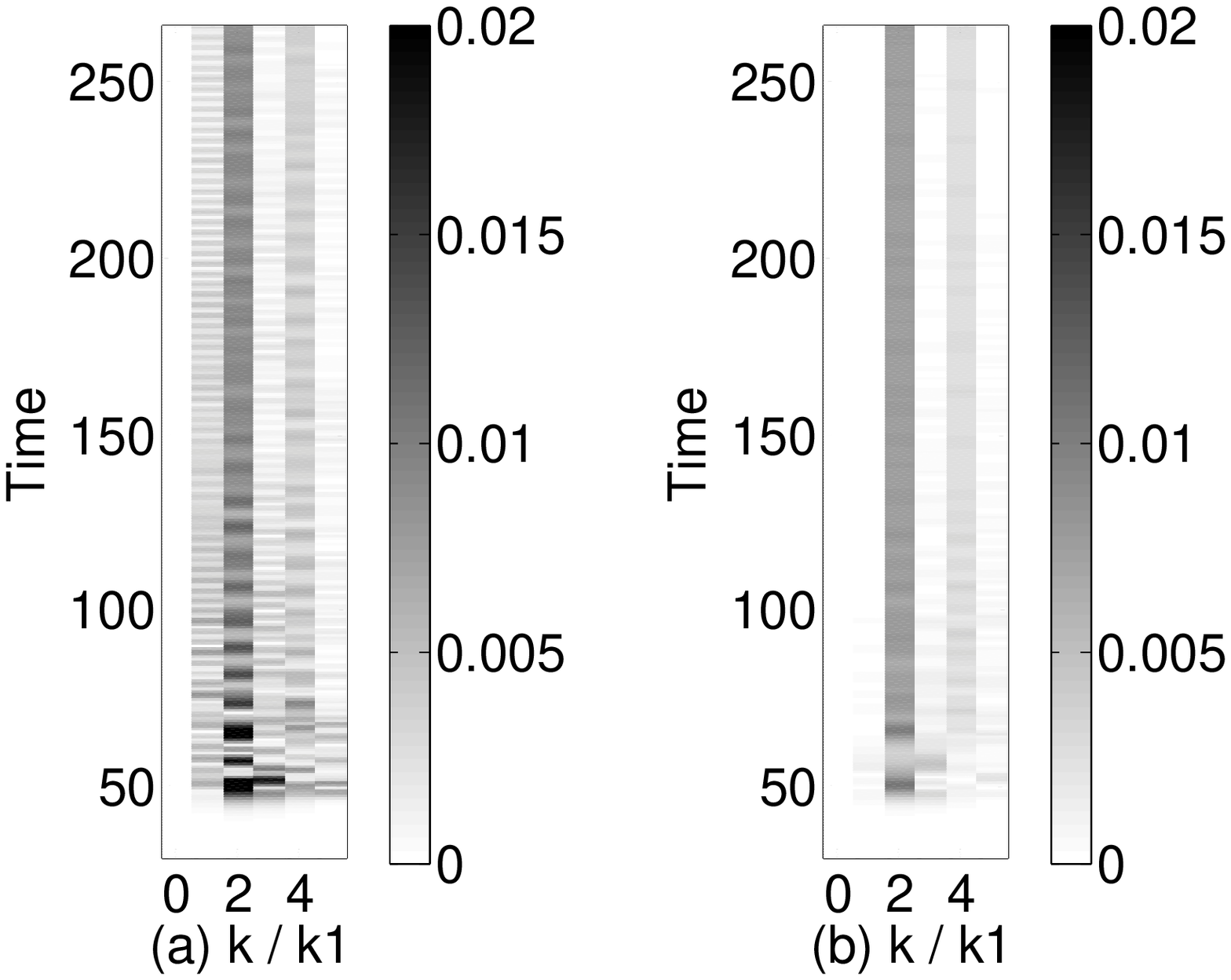}
\includegraphics[width=0.49\columnwidth]{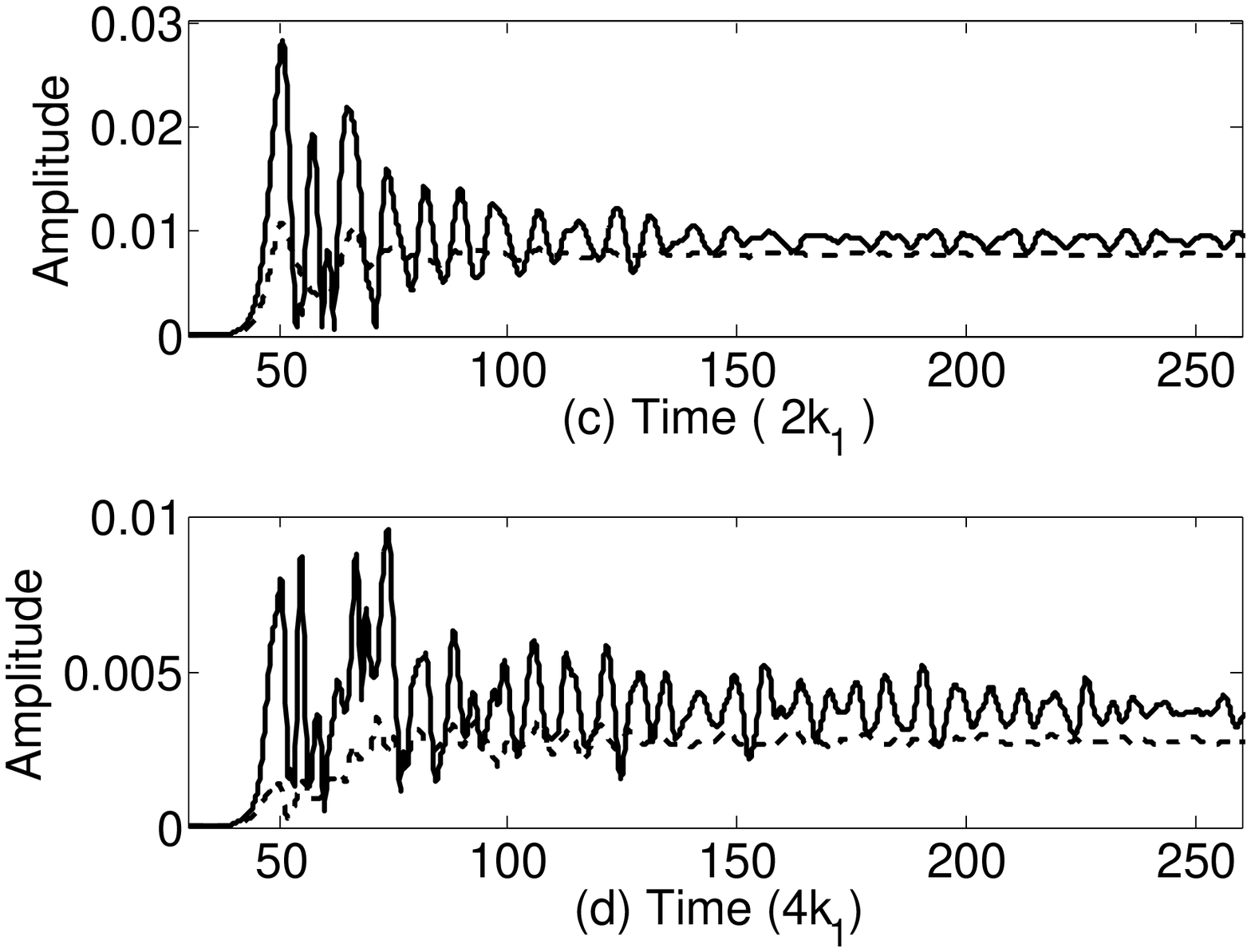}
\caption{The relevant part of the amplitude spectrum $E_x(k,t)$ is displayed 
for low $k$ in (a) and (b) shows that of $E_B(k,t)$. The wavenumbers are 
expressed in units of $k_1$. The amplitude moduli of the dominant modes are 
displayed for $k=2k_1$ in (c) and its first harmonic with $k=4k_1$ in (d), 
where the dashed curves correspond to $E_B$.}\label{Mfg4}
\end{center}
\end{figure}

Figure \ref{Mfg4} compares in more detail the moduli of the amplitude spectra 
$E_x (k,t)$ and $E_B (k,t)$. 
The amplitudes of the strongest modes fulfill $E_x (2k_1,t) \approx 2 E_B 
(2k_1,t)$ until $t=50$ (See also Fig. \ref{Mfg2}). The $E_x (2k_1,t)$ thus
overshoots $E_B (2k_1,t)$ and it oscillates around it after $t=50$. The
oscillation is damped and the amplitudes of $E_x (2k_1,t)$ and $E_B(2k_1,t)$ 
converge. The full spectra $E_x (k,t)$ and $E_B (k,t)$ reveal that the mode 
$k=4k_1$ is also important for $t>100$. It probably is a harmonic of the 
mode with $k = 2k_1$ and not an independently growing FI mode. Otherwise we 
would expect that the mode with $k \approx 3k_1$ also grows. The amplitude 
of $E_x (4k_1,t)$ is close to that of $E_B (4k_1,t)$.   

A dissipation mechanism for the interplaying $J_x$ and $E_x$ is present, 
which causes the damping and the convergence of $E_x (x,t)$ to $E_B(x,t)$. 
The damping persists after $t=65$, when $B_y$ is quasi-stationary in its 
moving reference frame. The term $B_y \partial_t E_z$ in (\ref{1D}) could, 
in principle, be one dissipation mechanism. However, even at $t\approx 50$ 
when $\partial_t E_z$ is largest and $B_y$ has developed in full, this term 
is weaker by more than one order of magnitude than the MPGF and the term 
$n_j E_x$ in simulation 1 (not shown). If the term $B_y \partial E_z$ would 
be the damping mechanism, this should have resulted in a noticable field 
damping also in the short simulation box with length $L_c$. A damping of 
$E_x (x,t)$ has not been observed in Ref. \cite{NewPoP}. The thermal pressure 
gradient force may provide this damping and we examine now the electron phase 
space density distribution.

\begin{figure}
\begin{center}
\includegraphics[width=0.49\columnwidth]{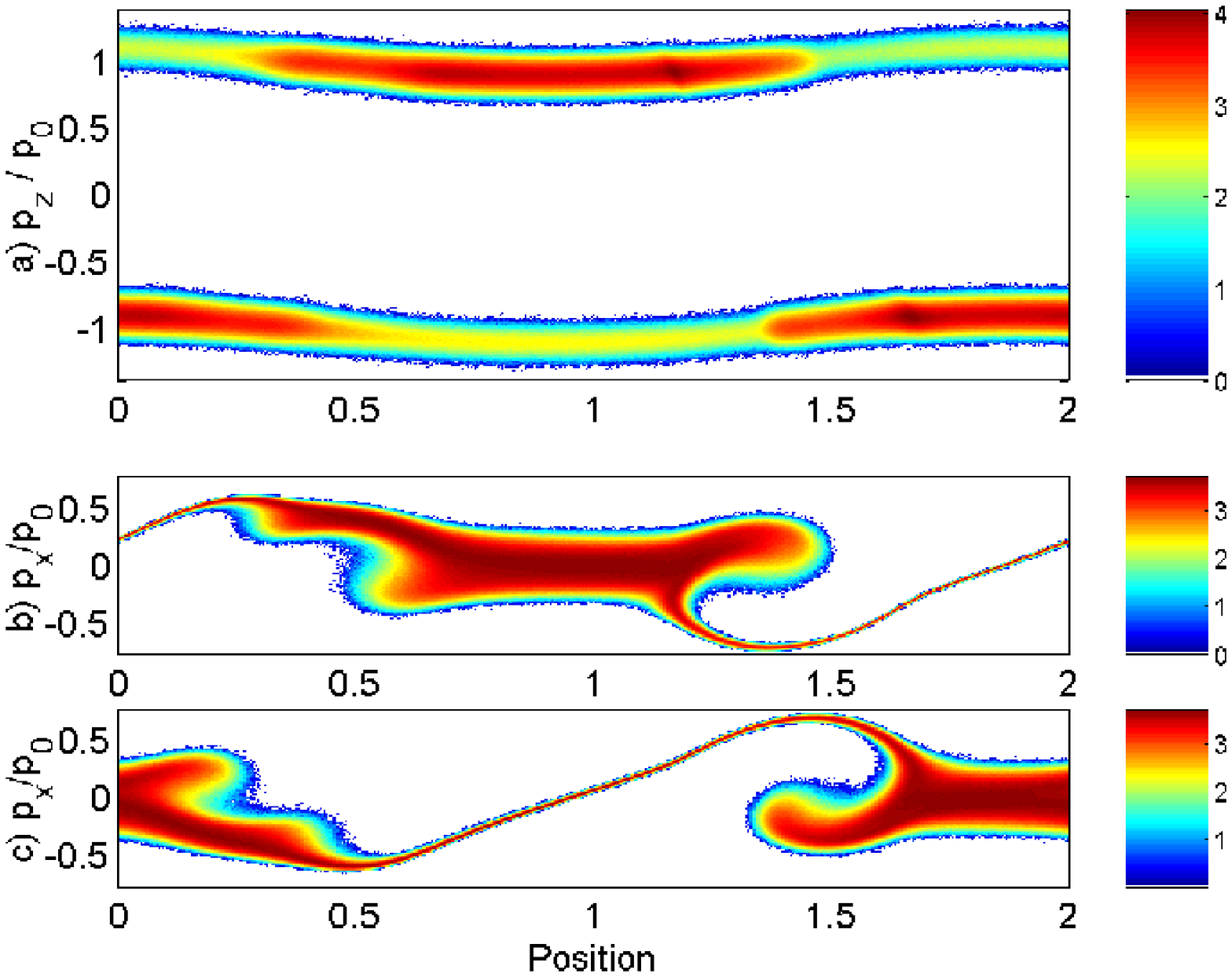}
\includegraphics[width=0.49\columnwidth]{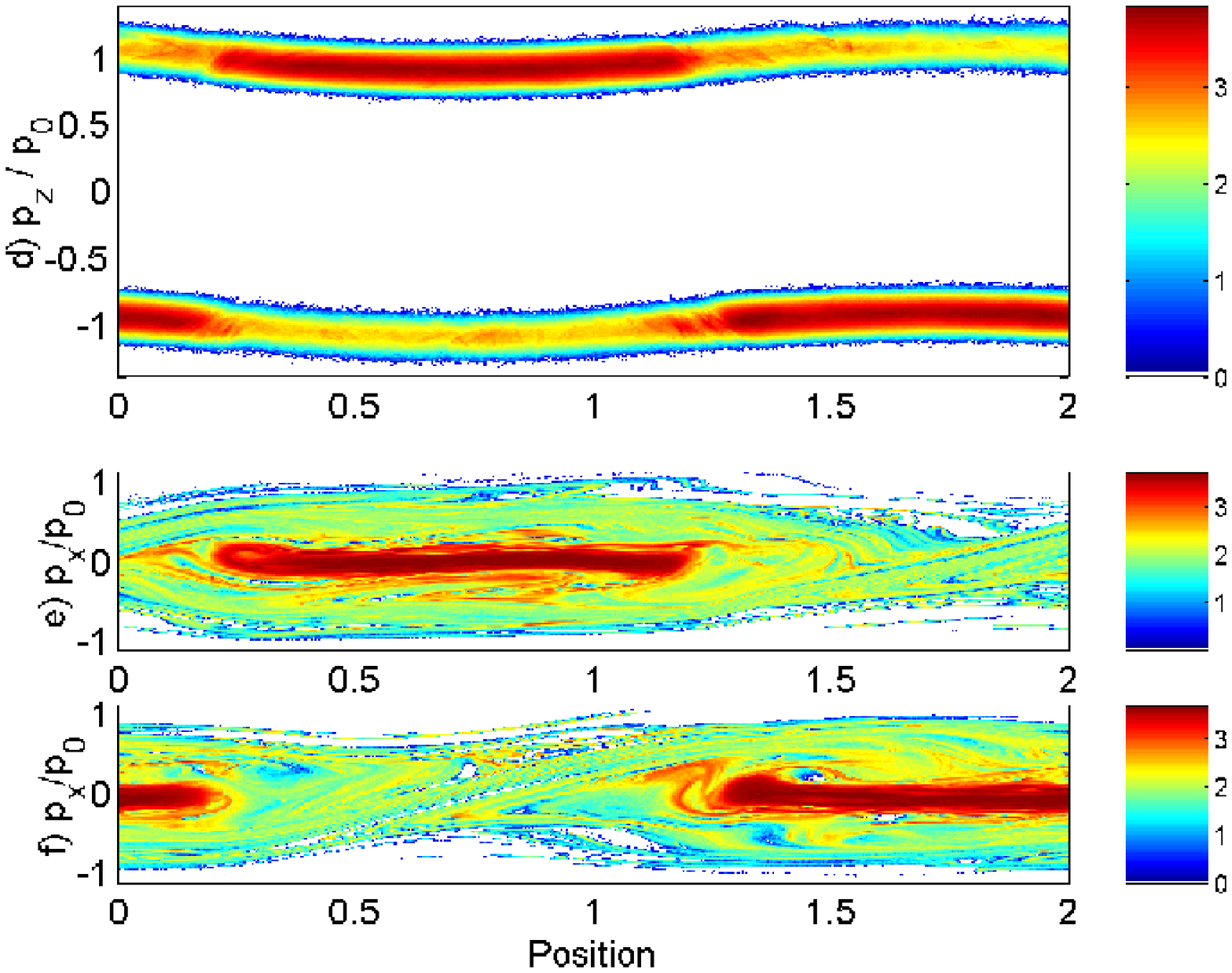}
\caption{(Colour online) The 10-logarithm of the phase space densities in 
units of CPs at the time $t=50$ (a-c) and $t=120$ (d-f) in the box $L_1$: 
Panels (a,d) show the total phase space density $f(x,p_z)$ with the beam 
momentum $p_0 = m_e v_b \Gamma (v_b)$. The phase space density $f_1(x,p_x)$ 
of beam 1 is shown in (b,e) and the $f_2(x,p_x)$ of beam 2 in (c,f). Both 
beams are spatially separated and (e,f) reveal cool electron clouds immersed 
in a hot electron background with momenta of up to $\approx p_0$.\label{Mfg5}}
\end{center}
\end{figure}

Figure \ref{Mfg5} displays the phase space distributions $f_1(x,p_x)$ 
of the beam 1 and the $f_2 (x,p_x)$ of the beam 2 at the times $t=50$ 
and $t=120$. The total phase space density $f(x,p_z)$ is shown for the 
same times. 
The beams reveal a high degree of symmetry already at $t=50$ and the filament 
centres are shifted along $x$ by $L_1 / 2$. The phase space structures in Fig. 
\ref{Mfg5}(b,c) are, however, different at the filament boundaries $x \approx 
0.5$ and $x\approx 1.5$. This difference is responsible for the deviation of 
the initial $B_y (x,t)$ from a sine curve in Fig. \ref{Mfg3}(a). The phase 
space distribution at late times reveals, that the electrons are heated along 
$p_x$ but not along $p_z$. The filament drift to lower $x$ is visible from 
Figs. \ref{Mfg5}(a,d) and agrees with the observed one of $B_y(x,t)$ in Fig. 
\ref{Mfg3}. The electrons are accelerated along $x$ to a peak speed $\sim
v_b$, which is more than twice that observed in the box with the length
$L_c$ \cite{NewPoP}. The peak electron kinetic energy due to the velocity
component along $x$ thus increases by a factor, which is comparable to the 
increase in the electrostatic potential as we go from a box with length 
$L_c$ to one with $L_1$. This strong electron heating is likely to result 
in higher thermal pressure gradient forces. The expression $\rmd_x n_1(x) 
\int v_x f(x,v_x)dv_x$ has been evaluated (not shown) at $t=75$ and its 
peaks reach values $\approx 0.1$, which are comparable to the MPGF. The
width of these peaks is small compared to the electron skin depth.

Movie 1 animates in time the 10-logarithmic phase space distributions 
$f_1(x,p_x)$ and $f_1(x,p_z)$ of the beam 1 in the simulation 1. The 
formation of the filaments is demonstrated. We observe a dense untrapped 
electron component immersed in an electron cloud that has been heated 
along the simulation direction by the saturation of the FI. The spatial 
width of the plasmon containing the dense bulk of the confined
electrons in $f_1(x,p_x)$ oscillates in time. The overlap of the filaments 
in Fig. \ref{Mfg5}(e,f) is thus time dependent and related through its 
current $J_x(x,t)$ to the oscillating $E_x (x,t)$ in Fig. \ref{Mfg3}(b). 
The phase space distribution $f_1(x,p_x)$ reveals small-scale structures 
(phase space holes) that gyrate around the centre of the filament. These 
coherent structures result in jumps in the thermal pressure.

\subsection{Simulation 2: Box length $L_2 = 2.8$}

\begin{figure}
\begin{center}
\includegraphics[width=0.49\columnwidth]{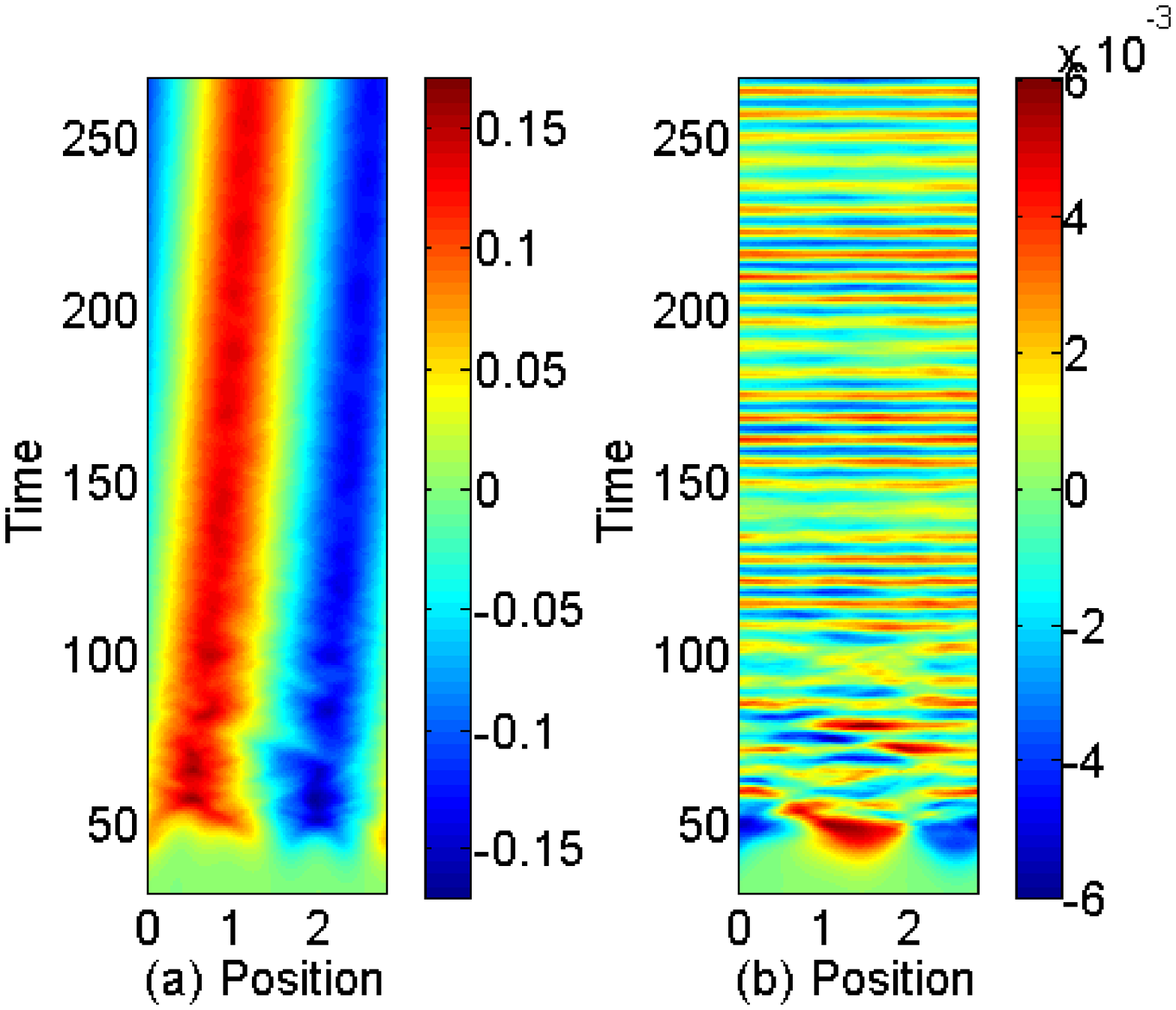}
\includegraphics[width=0.49\columnwidth]{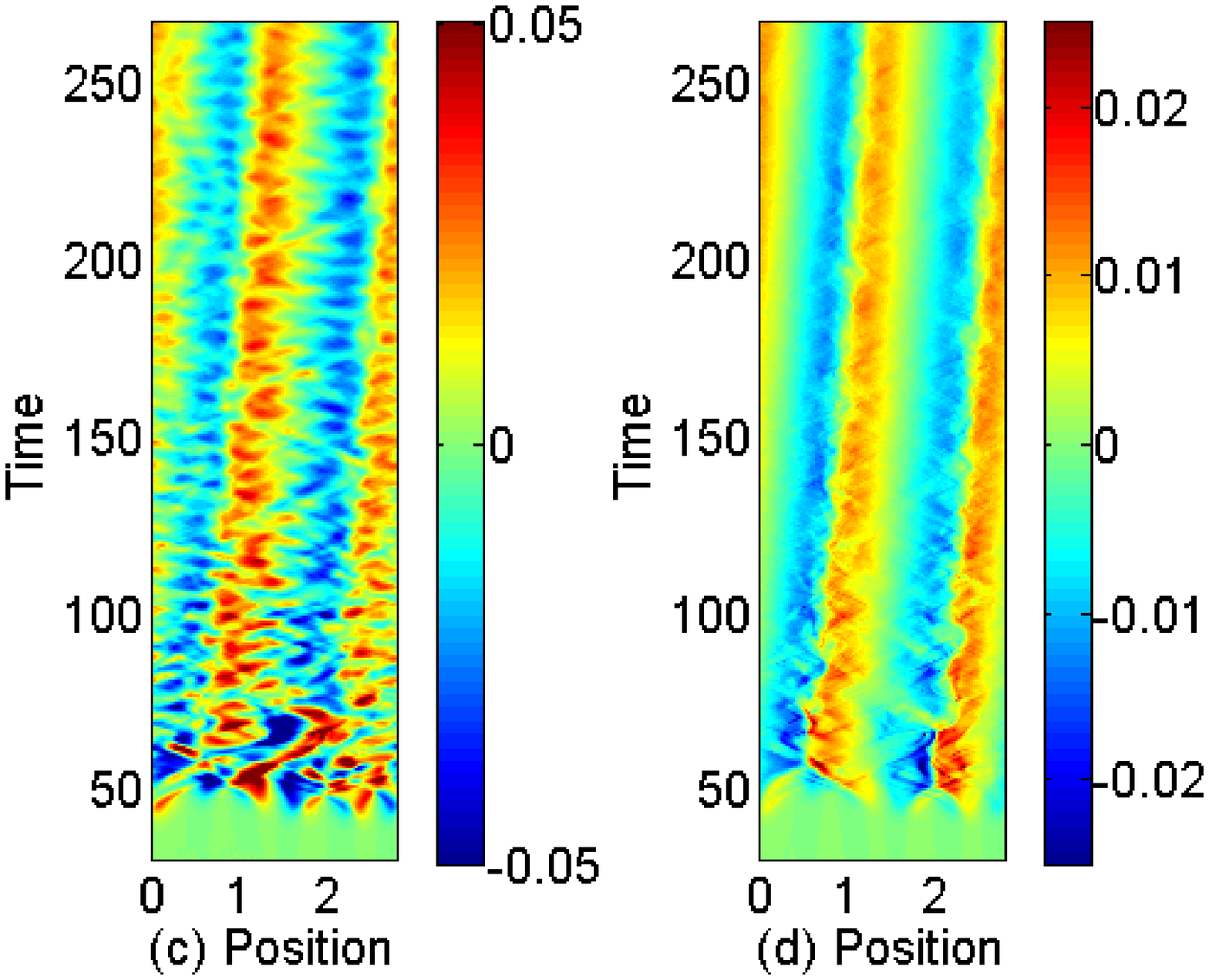}
\caption{The field amplitudes in the box $L_2$: The panels (a-d) show
$B_y$, $E_z$, $E_x$ and $E_B = -B_y d_x B_y$, respectively. The amplitude
of $B_y$ reaches a steady state value, which convects to increasing $x$ 
at a speed $< 0.01$. The $E_z$ and $E_x$ components are oscillatory 
in space and in time. The $E_z$ is phase-shifted by $90^\circ$ relative 
to $B_y$ when the fields saturate at $t\approx 50$. The $E_x$ and the $E_B$ 
are co-moving and $E_x(x)$ oscillates in time around a mean amplitude 
comparable to $E_B$ for $t > 100$.\label{Mfg6}}
\end{center}
\end{figure}

Figure \ref{Mfg6} displays the fields that grow in the simulation with
the box length $L_2=2.8$.
The growth rate map in Fig. \ref{Mfg1} demonstrates that the FI can 
drive simultaneously several modes in the simulation box. The mode with
$k_2=2\pi/L_2 \approx 2.25$ has, for example, a lower growth rate than 
that with $k \approx 2k_2$. We observe consequently oscillations in 
$B_y(x,t)$ along $x$, which are a superposition of several waves with a 
$k \ge k_2$ during the initial growth phase $40<t<50$. These oscillations 
merge and only one spatial oscillation of $B_y (x,t)$ and, thus, a single 
pair of filaments survive after the saturation at $t\approx 50$. The magnetic 
field structure convects to increasing values of $x$ at a speed $< 0.01$, 
but it is stationary in its rest frame after $t \approx 70$. The phase of 
$E_z(x,t)$ is shifted by $90^\circ$ with respect to $B_y(x,t)$ for $40 < t 
< 60$. The oscillations of $E_z(x,t)$ undergo a mode conversion during 
$60 < t < 100$ and we observe undamped oscillations with $k=0$ for $t>100$. 
The amplitude of these oscillations is modulated on a long timescale. The 
$E_x(x,t)$ and the $B_y(x,t)$ show no correlation until $t \approx 70$. 
Thereafter the spatial amplitude of $E_x (x,t)$ oscillates in time around 
$E_B(x,t)$. The force on an electron imposed by $E_x (x,t)$ is comparable 
to that imposed by $v_b B_y(x,t)$.

\begin{figure}
\begin{center}
\includegraphics[width=0.49\columnwidth]{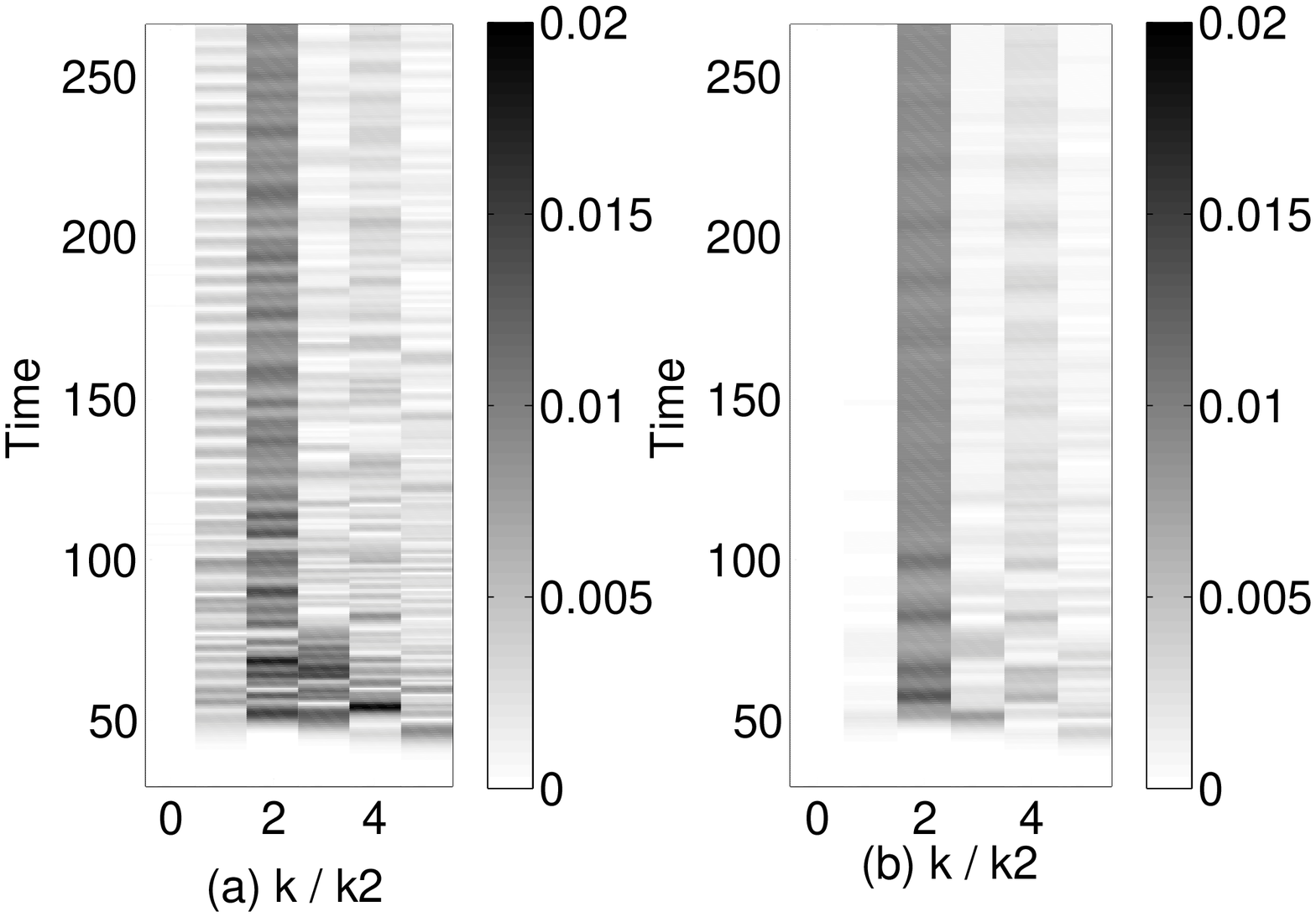}
\includegraphics[width=0.49\columnwidth]{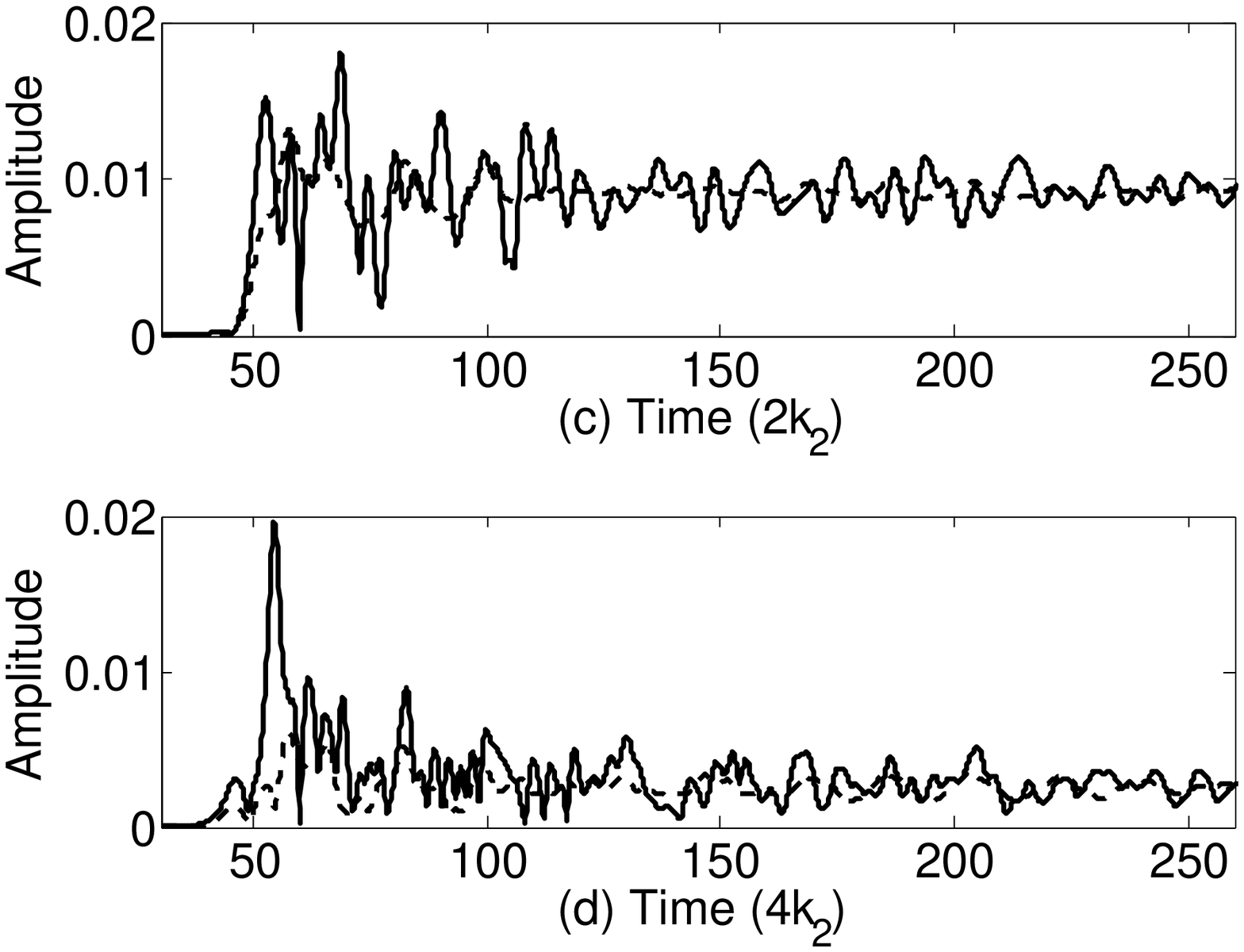}
\caption{The relevant part of the amplitude spectrum $E_x(k,t)$ is displayed 
for low $k$ in (a) and (b) shows that of $E_B(k,t)$. The wavenumbers are 
expressed in units of $k_2$. The amplitude moduli of the dominant modes are 
displayed for $k=2k_2$ in (c) and its first harmonic with $k=4k_2$ in (d), 
where the dashed curves correspond to $E_B$.}\label{Mfg7}
\end{center}
\end{figure}

A more accurate comparison of $E_x(x,t)$ and $E_B(x,t)$ is again provided 
by the moduli of their spatial amplitude (Fourier) spectra, $E_x (k,t)$ 
and $E_B(k,t)$. Figure \ref{Mfg7} displays $E_x(k,t)$ and $E_B(k,t)$ and 
compares in more detail $E_x(2k_2,t)$ with $E_B(2k_2,t)$ as well as
$E_x(4k_2,t)$ with $E_B(4k_2,t)$.
The amplitudes $E_x(2k_2,t) \approx 2E_B(2k_2,t)$ during the exponential 
growth phase of the FI for $25 < t < 45$ (See Fig. \ref{Mfg2}), the
amplitude moduli then have a local minimum and continue to grow after
this time. We identify the likely reason from $E_B(k,t)$ in Fig. 
\ref{Mfg7}(b). The $E_B(3k_2,t)$ competes with $E_B(2k_2,t)$ at $t \approx 
50$. 

A large amplitude modulus of $E_B(3k_2,t)$ evidences that $B_y(x,t)$ 
is not a sine wave at this time. If $B_y \propto \sin{(k_2x)}$, then $E_B 
\propto \sin{(k_2x)}\cos{(k_2x)}$ and $E_B(k,t)$ would be composed of a 
wave with $k=2k_2$. The periodic boundary conditions would also allow
for a $B_y \propto \sin{(2k_2,t)}$ and here $E_B$ would involve a wave
with $k=4k_2$. An $E_B(3k_2,t)$ can thus not be connected to a single
filamentation mode. During the linear growth phase of the FI prior to 
$t\approx 40$, the $J_z(x,t)$ can form structures with a wideband wavenumber 
spectrum (See Fig. \ref{Mfg1}) and their associated $B_y$ can grow 
independently. 

Once the MPGF in Eq. \ref{1D} has reached a significant strength, the FI 
saturates. The strength of the MPGF increases with $k$, due to the larger
$\rmd_x B_y(x,t)$ of the rapid oscillations. The $B_y \propto \sin{(k_2x)}$ 
should maximize the magnetic field strength for a given MPGF. This may
explain why this mode is the dominant one after $t=70$ despite its lower 
growth rate. The decrease of $E_B(2k_2,t)$ in Fig. \ref{Mfg2}(d) at 
$t\approx 45$ is tied to the saturation of $E_B(3k_2,t)$. The $E_x(k,t)$ 
in Fig. \ref{Mfg7}(a) has a broadband spectrum within $50<t<75$, which is 
probably caused by the current $J_x$ arising from the rearrangement of the 
filaments. After this time, $E_x(2k_2,t)\approx E_B(2k_2,t)$ and 
$E_x(4k_2,t)\approx E_B(4k_2,t)$. The $E_x(2k_2,t)$ does not show 
oscillations around $E_B(2k_2,t)$ as the simulation 1. The filament 
rearrangement inhibits an oscillatory equilibrium between $J_x$ and $E_x$.

\begin{figure}
\begin{center}
\includegraphics[width=0.49\columnwidth]{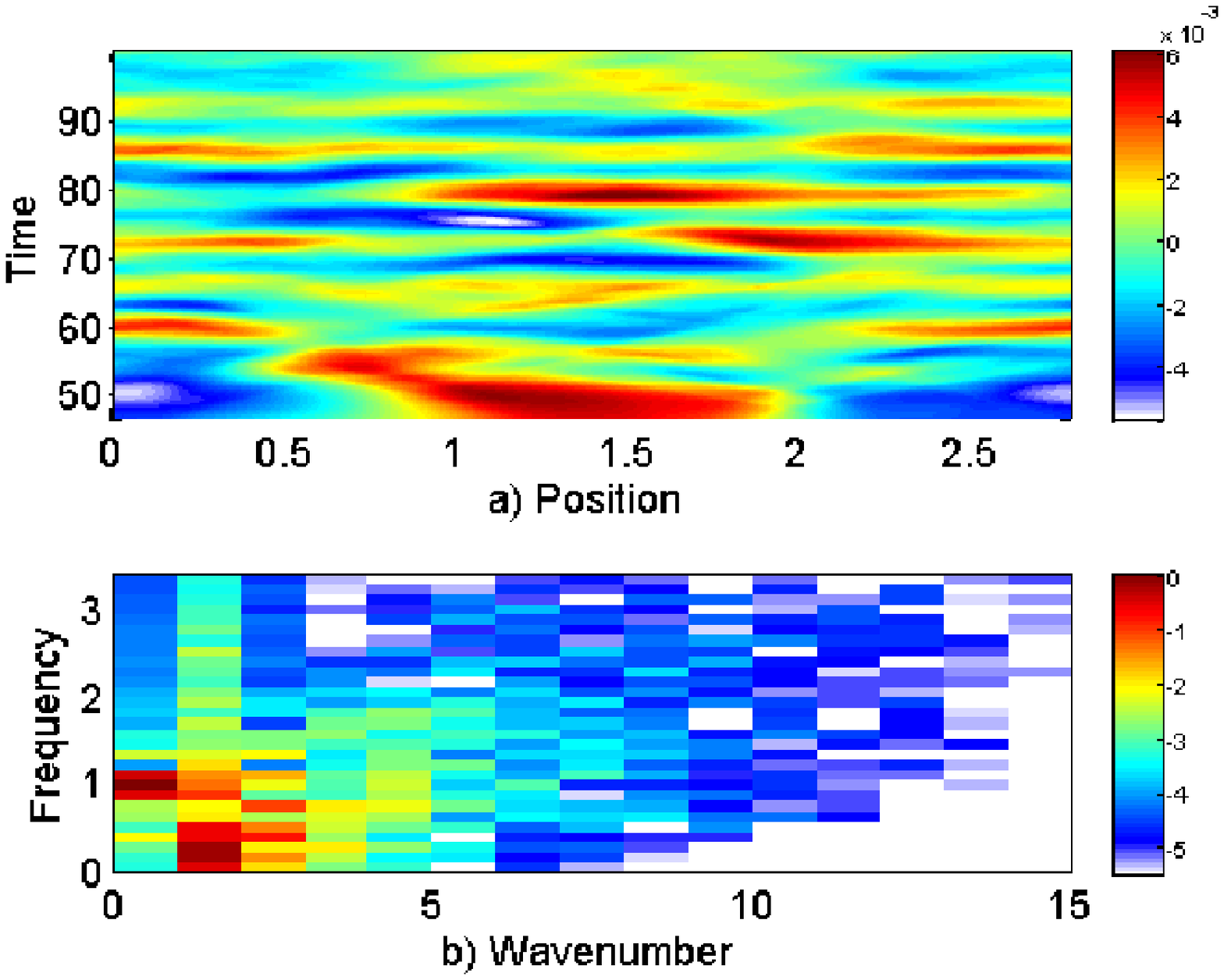}
\includegraphics[width=0.49\columnwidth]{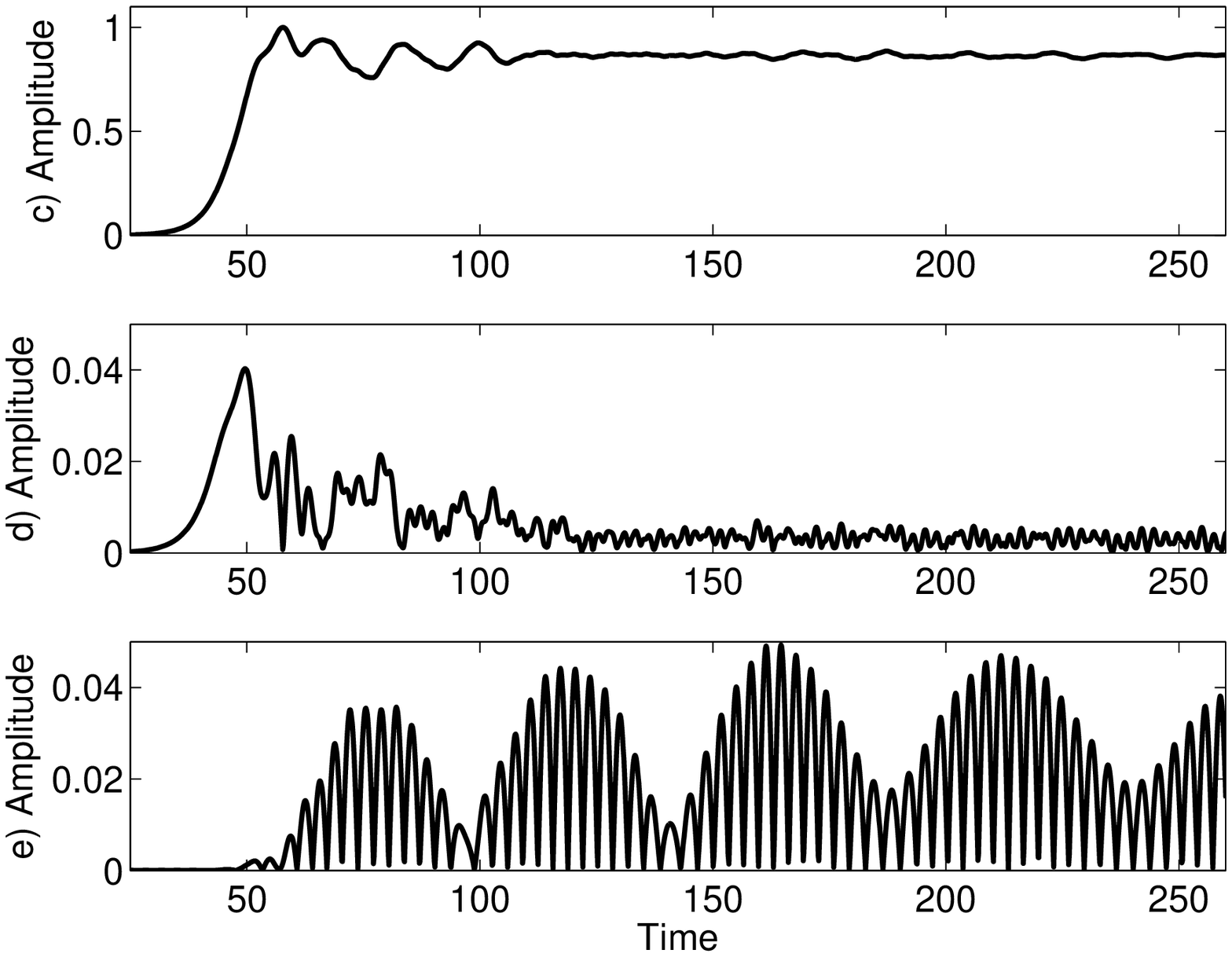}
\caption{(Colour online) A time-interval of $E_z (x,t)$ and the 10-logarithm 
of its power spectrum $P_{EZ} (k,\omega)$ are displayed in (a) and (b). 
Wavenumbers are given in units of $k_2$. Peak 1 is at $\omega < 0.5$ and 
$k=k_2$. Peak 2 is observed at $k=k_2$ and $\omega \approx 1$ and peak 3 at 
$k=0$ and $\omega \approx 1$. The $B_y (k_2,t)$ is shown in (c), the 
$E_z(k_2,t)$ in (d) and $E_z(0,t)$ in (e), all normalized to the maximum of 
$B_y(k_2,t)$.}\label{Mfg8}
\end{center}
\end{figure}

Figure \ref{Mfg8} examines the mode conversion of the electromagnetic $E_z$ 
component observed in Fig. \ref{Mfg6}(b). The $P_{EZ} (k,\omega)$ is the 
squared modulus of the Fourier transform of $E_z (x,t)$ over space and over 
$45 < t < 100$. 
The dispersion relation shows three peaks. Peak 1 has a $k=k_2$ and $\omega 
< 0.5$ and it is tied to the $E_z (x,t)$ of the FI mode. This mode grows 
exponentially and aperiodically. Its frequency spectrum is thus spread out 
along $\omega$. Its energy can leak into the peak 2 at $k=k_2$ and $\omega 
\approx 1$. The $E_z(x,t)$ is orthogonal to $B_y(x,t)$ and peak 2 corresponds 
to an extraordinary mode, similar to the slow extraordinary mode. Peak 3 has 
a $k=0$ and $\omega \approx 1$ and it corresponds to a spatially uniform 
oscillation in an extraordinary mode branch. The intermittent behaviour of 
$E_z(x,t)$ in Fig. \ref{Mfg8}(a) results in a broadband spectrum in $k$ and 
$\omega$. These turbulent wave fields can couple energy directly to the 
high-frequency electromagnetic modes and excite a discrete spectrum if 
the boundary conditions are periodic \cite{Califano}. 

The interplay of the waves belonging to the three peaks in Fig. \ref{Mfg8}(a) 
is assessed with the moduli of the amplitude spectra $B_y(k_2,t)$, $E_z(k_2,t)$ 
and $E_z(0,t)$ in Figs. \ref{Mfg8}(c-e). The $B_y(k_2,t)$ and $E_z(k_2,t)$ 
grow at the same exponential rate until they saturate at $t\approx 50$,
evidencing that they belong to the same FI mode. The $B_y(k_2,t)$ maintains 
its amplitude after $t=50$, while $E_z(k_2,t)$ decreases until $t \approx 120$ 
and remains constant thereafter. The $E_z(0,t)$ grows in the same time 
interval to its peak amplitude, which suggests a parametric interaction 
between these modes. The amplitude modulation in Fig. \ref{Mfg8}(e) must 
be caused by a beat between two waves, which are similar to the 
slow- and fast extraordinary modes in the limit $k=0$. Both modes are 
undamped on the resolved timescales. One may interpret the parametric
interaction as a three-wave coupling between the waves corresponding to the
peaks 1-3 in Fig. \ref{Mfg8}(b), resembling the system of Ref. \cite{Sharma}. 
However, here the $B_y(x,t)$ varies spatially and the parametric interaction 
may involve more of the waves of the spectrum in Fig. \ref{Mfg8}(b).  

\begin{figure}
\begin{center}
\includegraphics[width=0.49\columnwidth]{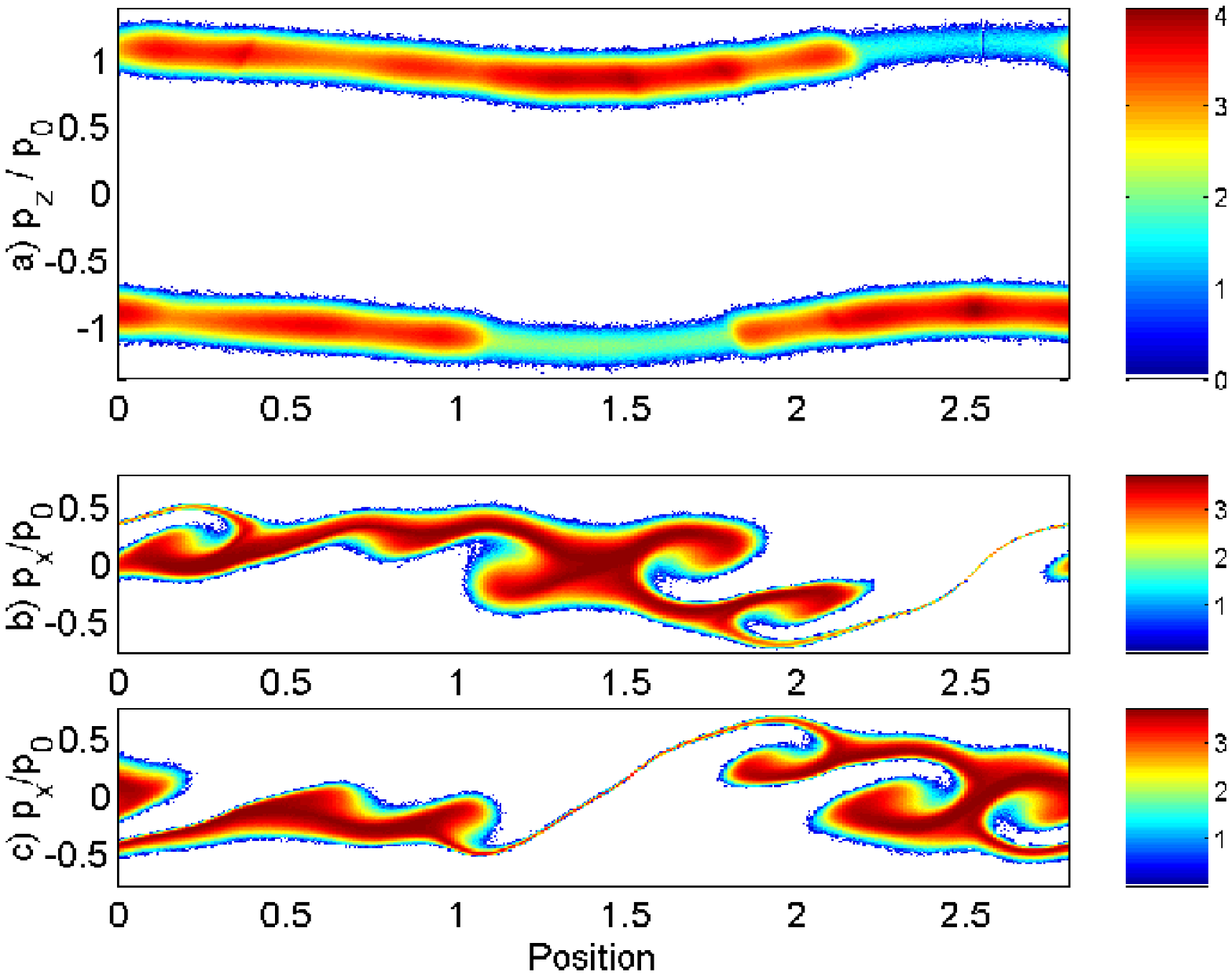}
\includegraphics[width=0.49\columnwidth]{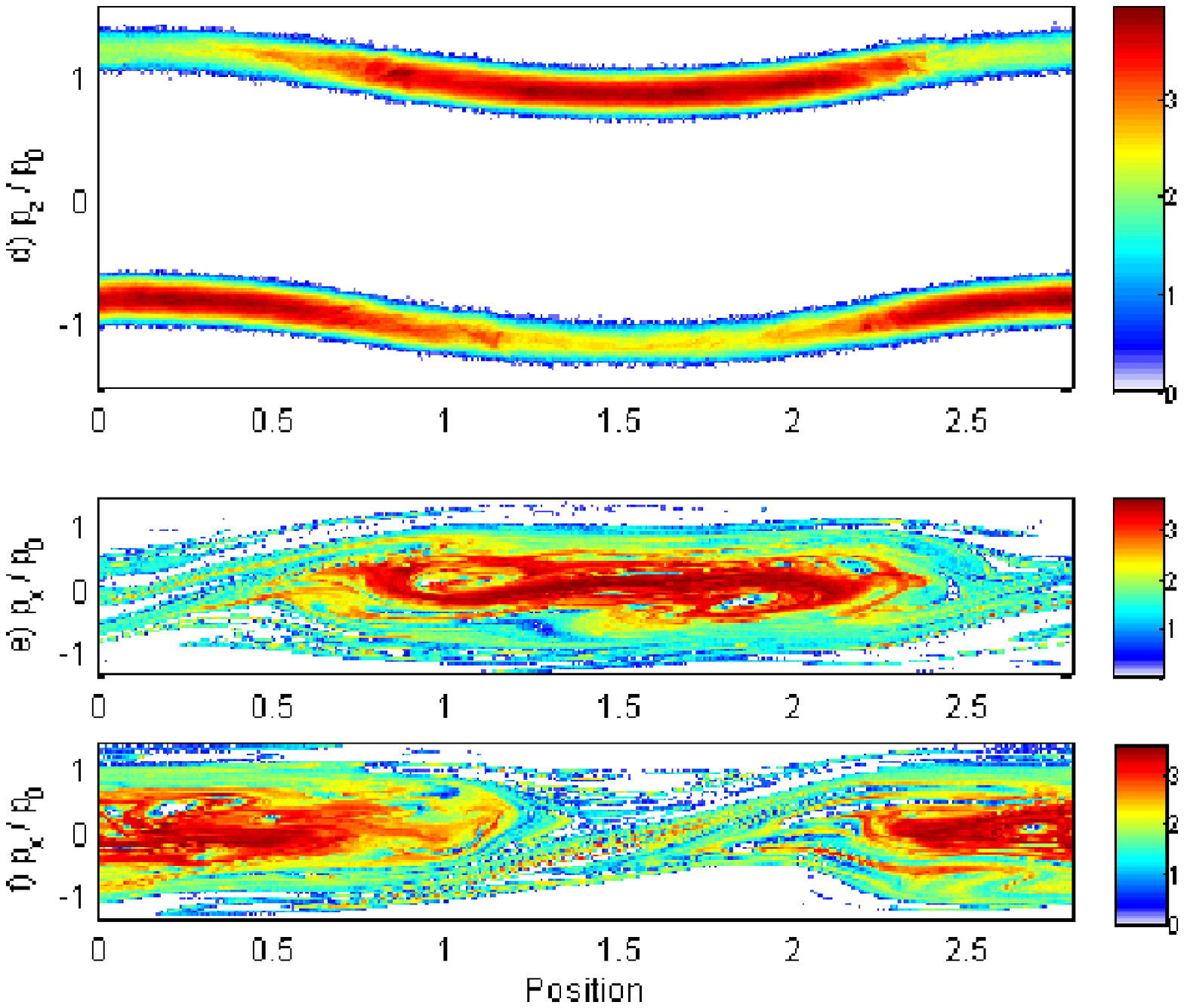}
\caption{The 10-logarithmic phase space densities in units of CPs at $t=50$ 
(a-c) and $t=120$ (d-f) in the box $L_2$: Panels (a,d) show the total
distribution $f(x,p_z)$ with $p_0 = m_e v_b \Gamma (v_b)$. The beam 
temperature along $p_z$ is unchanged. The distribution $f_1(x,p_x)$ of beam
1 is shown in (b,e) and the $f_2(x,p_x)$ of beam 2 in (c,f). The electrons 
of both beams spatially separate and (e,f) reveal a dense electron component
immersed in a tenuous hot electron background, which reaches a thermal 
width $\approx p_0$.\label{Mfg9}}
\end{center}
\end{figure}
Figure \ref{Mfg9} displays the phase space densities $f_{1,2}(x,p_x)$ and 
$f(x,p_z)$ at the times $t=50$ and $t=120$. Figure \ref{Mfg9}(a) demonstrates 
that the electrons of both beams have been rearranged by the FI. The filaments 
have not yet reached the stable symmetric configuration, because the most 
pronounced density minima at $x\approx 1.5$ for beam 2 and at $x\approx 2.5$ 
for beam 1 are not shifted by $L_2/2$. This asymmetry results in the 
$E_B (3k_2,t)\neq 0$ and in the broadband $E_x(k,t)$ at this time in Fig.
\ref{Mfg7}. The spatial gradients of $B_y(x,t)$ and $E_x (x,t)$ are high at 
$t\approx 50$ and the Lorentz force changes rapidly with $x$, explaining the 
complex phase space structuring in Fig. \ref{Mfg9}(b,c). The phase shift of 
$L_2/2$ of the density maxima of both beams has been reached at $t=120$ in 
Fig. \ref{Mfg9}(d). The $B_y(x,t)$ is stationary in its rest frame at this 
time in Fig. \ref{Mfg6}(a). The electrons are heated up from an initial 
thermal spread of $p_x / p_0 \approx 0.05$ with $p_0 = m_e v_b \Gamma (v_b)$ 
to a peak value of $p_x \approx p_0$ in Fig. \ref{Mfg9}(e,f). The mean 
momentum of each beam varies along $p_z$ in response to a drift imposed by 
$E_x(x,t)$ and $B_y(x,t)$ but no heating is observed in this direction.

Movie 2 shows the 10-logarithmic phase space density projections $f_1(x,p_x,t)$ 
and $f_1(x,p_z,t)$ of beam 1 in the simulation 2. It demonstrates that only 
the core electrons in Fig. \ref{Mfg9} remain spatially confined. The heated 
electrons, which have in some cases reached a momentum $p_x$ that is comparable 
to the initial beam momentum, are untrapped. The heated electrons move 
practically freely and they ensure that the beam confinement is not perfect. 
The trapped electrons maintain the $J_z (x,t) \neq 0$ and, thus, the 
$B_y(x,t)\neq 0$. The trapped electrons slowly move to larger values of $x$. 
The associated shift of $J_z(x,t)$ causes the slow drift of $B_y(x,t)$ in 
Fig. \ref{Mfg6}(a). The movie visualizes the formation of the phase space 
beams and their evolution into phase space holes in $f_1 (x,p_x)$.

\section{Discussion}

We have examined here the electron beam filamentation instability (FI) in 
one dimension and in an initially unmagnetized plasma with immobile ions. 
The FI has been driven by nonrelativistic symmetric electron beams with 
the same initial conditions as those considered previously 
\cite{Rowlands,NewPoP}. The electric field along the one-dimensional
box, which is oriented orthogonally to the beam velocity vector, can only
be generated nonlinearly if the beams are symmetric \cite{Tzoufras}. The
fluid equations show that the relevant nonlinear mechanisms can be the 
magnetic pressure gradient force (MPGF), the thermal pressure gradient 
force and a term due to the displacement current. The magnetic tension 
may become important in multi-dimensional simulations, but not for initial 
conditions similar to ours \cite{EPS}. The term due to the displacement
current is weak in our simulations.

It has been observed in Ref. \cite{NewPoP} that the electrostatic field
performs after the saturation of the FI undamped oscillations around a 
time-stationary background electric field. The amplitude of the oscillatory 
and of the background electric field are both given by $E_B (x,t) \approx 
-B_yd_x B_y$. The phases of both fields are fixed such, that $E_x (x,t_0)=
2E_B(x,t_0)$ at the saturation time $t_0$. This amplitude ensures that the 
nonlinear terms due to the MPGF and due to the electrostatic field cancel 
each other approximately in the fluid equations when the FI saturates. The 
thermal pressure gradient force did not visibly contribute in the simulation 
of the small filament pair \cite{EPS}, possibly because of the only modest 
heating of the initially cool beams. Here we have assessed the importance of 
the filament size with the help of two 1D PIC simulations, which used two 
different box lengths that were larger than that of the 1D box in Ref. 
\cite{NewPoP,EPS}. The initial conditions for the plasma were otherwise 
identical. 

We summarize our findings as follows. We have demonstrated for both 
simulations, that $E_x (x,t\le t_0) \approx 2E_B(x,t \le t_0)$ during the 
full exponential growth phase and not just at the saturation time $t_0$. 
The FI thus adjusts the electrostatic field during its exponential growth 
phase such, that the dominant nonlinear terms cancel each other. Magnetic 
trapping states that the FI saturates, when the magnetic bouncing frequency 
is comparable to the linear growth rate. The exponential growth rates for the 
two simulations considered here and that in Ref. \cite{NewPoP} are close. 
The amplitude reached by the magnetic field prior to its satuation thus 
increases with the box length. We found that the electrostatic potential 
driven by the MPGF is 5-6 times stronger for the box sizes used here than 
for the short box in Ref. \cite{NewPoP}, while the initial mean kinetic 
energy of the electrons is the same. Consequently, the electron heating 
is stronger and the plasma processes more violent for large filaments. 
Magnetic trapping is, however, not the exclusive saturation mechanism. 
The electrostatic forces are comparable in strength to the magnetic forces 
when the FI saturates \cite{Pukhov2,Califano}.
 
The electrostatic field during the intermittent phase has differed in our 
two simulations from that observed in Ref. \cite{NewPoP}. The movies 
demonstrated that this phase involves the formation of large nonlinear 
structures (phase space holes) in the electron distribution, which can 
result in steep gradients of the thermal pressure and in the generation 
of solitary (bipolar) electrostatic wave structures that are independent 
of the fields produced by the FI. The thermal pressure gradient force is 
comparable to that of the other nonlinear terms, but only over limited 
spatial intervals. The electric field component along the beam velocity
vector has undergone a mode conversion. Its energy leaked into the 
high-frequency electromagnetic modes \cite{Califano}.

The wavenumber spectrum of the electrostatic field correlated well with 
that of the MPGF in simulation 1, but the peak electric field overshot 
the expected one. The electrostatic field performed damped oscillations 
around $E_B$ and both converged eventually to the same value. The wavenumber 
spectrum of the electrostatic field in simulation 2 deviated from that of the
MPGF in the intermittent phase. Its wavenumber spectrum was broadband, while 
that of the MPGF was quasi-monochromatic. The amplitude modulus of the 
electrostatic field at the wavenumber, which corresponds to the dominant 
Fourier component of the MPGF, jumped to the value expected from the MPGF. 
It did not overshoot and it was non-oscillatory. 

Both simulations here have evidenced that the magnetic field driven by 
the FI organized itself such, that we obtained one oscillation in the 
simulation box after the intermittent phase. This is remarkable, because 
the exponential growth rate of the fundamental wavenumber is below that 
of its first harmonic. Long waves excert a lower MPGF for a given amplitude 
and the dominance of the fundamental wavenumber may thus result from the 
lower nonlinear damping of this mode compared to that of its harmonics. The 
mode with the fundamental wavenumber considered in Ref. \cite{NewPoP} has a 
higher growth rate than its harmonics and the absent mode competition may 
have facilitated the undamped oscillations around the equilibrium. However, 
the amplitude of the electrostatic field in the two simulations discussed 
here eventually converged to that expected from the MPGF and $E_B$ is thus 
a robust estimate for the electrostatic field driven by the MPGF for the 
considered case. This robustness explains, why a connection between the 
electrostatic field and the MPGF has been observed in a 2D PIC simulation 
\cite{EPS}, where no equilibrium can be reached due to the filament mergers. 

This estimate does, however, not apply if positrons are present. Their 
current reduces that of the electrons. If equal amounts of electrons and 
positrons are present, the electrostatic field driven by the MPGF is 
suppressed alltogether \cite{Posi}. Mobile protons will react in particular 
to the stationary electric field \cite{Califano} and they will modify 
through their charge modulation the balance between the electrostatic field 
and the MPGF. Highly relativistic beam velocities will probably also modify 
the balance between the MPGF and the electron currents it drives. We leave 
relativistic beams to future work.

{\bf Acknowledgements} The authors acknowledge the support by 
Vetenskapsr\aa det and by the projects FTN 2006-05389 of the Spanish 
Ministerio de Educacion y Ciencia and PAI08-0182-3162 of the Consejeria 
de Educacion y Ciencia de la Junta de Comunidades de Castilla-La Mancha. 
The HPC2N has provided the computer time.

\end{document}